\documentclass[cameraready]{Interspeech}

\title{Probing and Mitigating Hallucinations in Speech-augmented Language Models for Automatic Speech Recognition via Small Language Models}

\author[affiliation={}]{Bi-Cheng}{Yan}
\author[affiliation={}]{Jhih-Rong}{Guo}
\author[affiliation={}]{Fu-An}{Chao}
\author[affiliation={}]{Berlin}{Chen}

\address{
    National Taiwan Normal University, Taiwan
}

\email{\{80847001s, jhihrong, fuanchao, berlin\}@ntnu.edu.tw}


\usepackage{comment}
\usepackage{bbm}
\usepackage{dsfont}
\usepackage[skip=4pt]{caption} 
\usepackage{adjustbox}
\usepackage{threeparttable}
\usepackage{booktabs} 
\usepackage{multirow}
\usepackage{graphicx}
\usepackage{xcolor}
\usepackage{soul}
\usepackage{makecell}
\usepackage{multirow}

\definecolor{asr-yellow}{rgb}{1.0, 0.945, 0.624}
\definecolor{asr-blue}{rgb}{0.878, 0.937, 1.0}
\newcommand{\hlyellow}[1]{{\sethlcolor{asr-yellow}\hl{#1}}}
\newcommand{\hlblue}[1]{{\sethlcolor{asr-blue}\hl{#1}}}

\begin{document}

\maketitle

\begin{abstract}
The widespread success of large language models in natural language processing has spurred the development of speech-augmented language models (SLMs) to extend their reasoning capabilities from textual inputs to spoken ones. Despite impressive performance, SLMs remain susceptible to hallucinations, posing challenges in practical utility. To this end, we first probe hallucinations in SLMs for automatic speech recognition (ASR) via the lens of causal mediation and behavioral analysis. To mitigate fabricated ASR hypotheses, we propose AudioSLM, a compact yet effective ASR framework built upon a small language model, which leverages alignment cues and cross-attention layers to synergistically bolster acoustic modeling and cross-modal interactions. Experiments on LibriSpeech reveal that hallucinations arise from multi-head self-attention modules with an excessive attention bias towards textual tokens. AudioSLM considerably eases ASR hallucinations and outperforms some LLM-based ASR models.
\end{abstract}

\section{Introduction}

As witnessed by the impressive success of large language models (LLMs) in the natural language processing community, a growing interest has surged in broadening LLM capabilities to process and understand speech signals \cite{touvron2023llama, han2024vall}. By coupling a speech encoder with a language model, speech-augmented language models (SLMs) have demonstrated promising performance in transducing complex acoustic patterns into contextually relevant outputs (e.g., transcriptions, captions, and intents), serving as a more natural and direct interface for human-AI interaction \cite{cui2025recent, peng2025vibevoice}. However, the phenomenon of hallucinations persists, as SLMs inadvertently generate content that diverges from the spoken inputs, posing an arduous challenge to the reliability and practical deployment of current SLMs.

Existing literature in natural language processing and computer vision suggests that hallucinations primarily arise from insufficient grounding on input modalities \cite{agarwal2020towards}. Early studies reveal that hallucinated words often emerge as backbone language models over-prioritize linguistic priors and rely heavily on previously generated tokens, sidelining the information provided by visual and textual modalities \cite{wu2022overcoming}. To curb the over-reliance on language priors, several methods have been proposed, which broadly fall into two active strands of research. The first line of research centers on post-processing strategies, such as post-hoc corrections \cite{huang2025survey, lee2024volcano} and specialized decoding algorithms \cite{kim2023exposing}. Another research direction aims to reduce the lexical co-occurrence bias inherent in backbone language models via data augmentation and preference optimization \cite{xie2024v, sun2024aligning, zhao2025beyond}.

\begin{figure}
    \centering
    \includegraphics[width=0.80\linewidth]{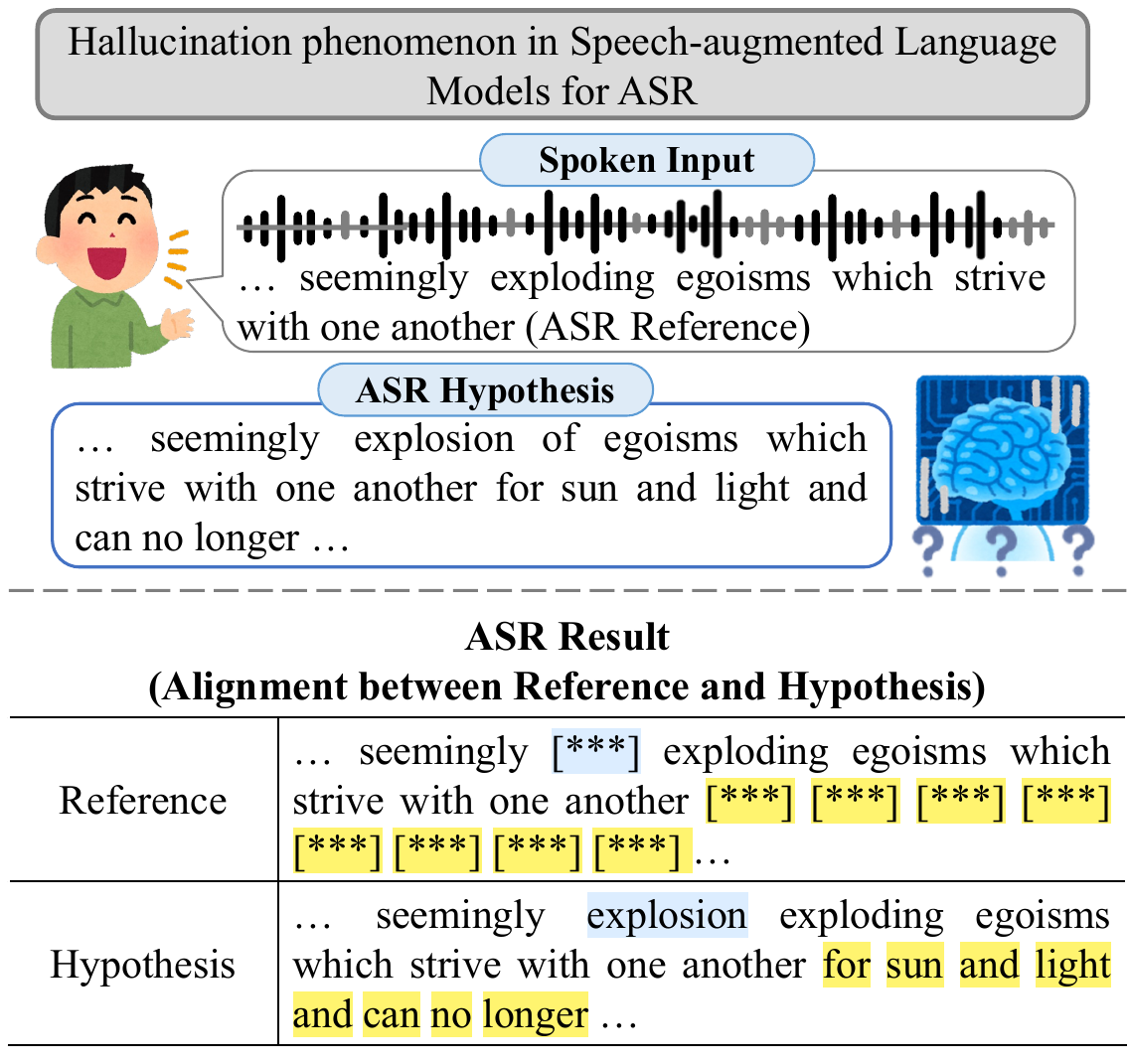}
    \caption{A running example illustrates ASR hallucination for an SLM. We define hallucinated words as a special case of insertion error, where fabricated words are highlighted in \hlyellow{yellow} and general standard inserted words are marked in \hlblue{blue}. }
    \label{fig:fig1}
\end{figure}
Despite significant strides in mitigating hallucinations for vision-language and text-only models, this issue remains relatively underexplored in SLMs. In response, this paper focuses exclusively on their impacts on automatic speech recognition (ASR), systematically investigating the underlying causes of hallucinations in SLMs. As illustrated in Figure \ref{fig:fig1}, we first characterize hallucinated words as a specific subtype of insertion error, referring to additional (or fabricated) tokens in the ASR hypothesis that lack corresponding references after the alignment process. Next, a series of analyses are conducted to uncover the causal contributions of internal components within SLM to fabricated token generation through the lens of attribution and intervention \cite{covert2021explaining}. Finally, we propose AudioSLM, a compact yet effective framework built upon a small language model that mitigates ASR hallucinations through two complementary strategies. Our method first leverages alignment features from a Connectionist Temporal Classification (CTC) module to provide reliable temporal cues for better acoustic modeling, and further enhances the backbone model with cross-attention layers to facilitate fine-grained alignment between acoustic and textual modalities. Through causal mediation analysis on the SLM components, we find that multi-head self-attention (MHA) modules contribute more significantly to hallucinations than multi-layer perceptron (MLP) modules. A closer inspection of the attention weights on hallucinated tokens reveals a pronounced tendency to over-prioritize textual context. Extensive experiments conducted on the LibriSpeech dataset with SmolLM2 model family \cite{allal2025smollm2} as the backbone demonstrate the effectiveness of our method in mitigating hallucinations while achieving competitive performance compared to some prior arts of LLM-based ASR methods.
\begin{figure}
    \centering
    \includegraphics[width=0.90\linewidth]{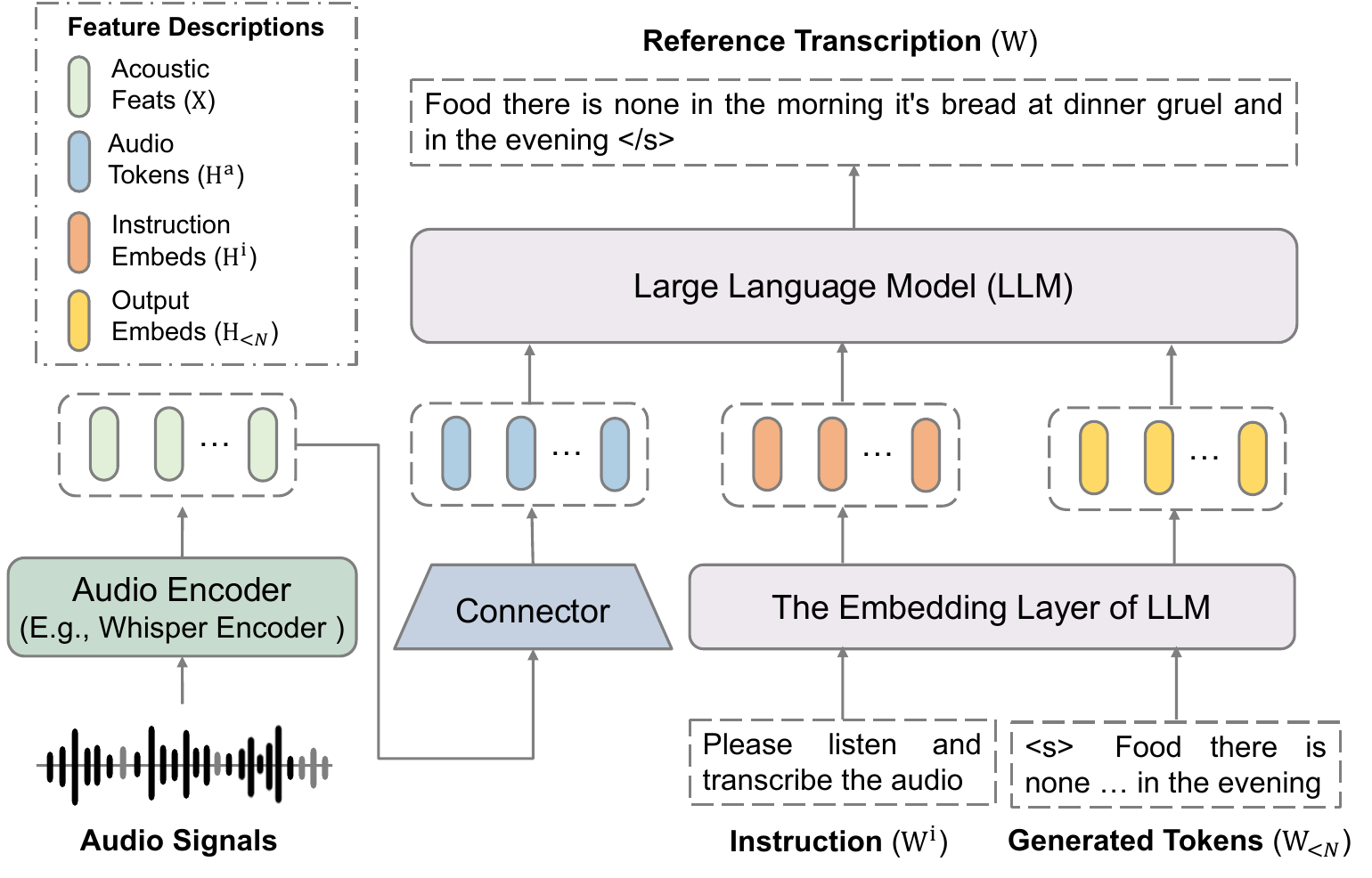}
    \caption{SLM architecture for ASR, comprising a speech encoder for acoustic feature extraction, a connector network for modality bridging, and an LLM for text generation.}
    \label{fig:fig2}
\end{figure}
\section{Preliminary and Problem Statement}
\textbf{Hallucination Error Rate for ASR.} We formally define the hallucination error rate (HER) as the ratio of the hallucinated tokens to the total number of reference (ground-truth) tokens: 
\begin{align}
  \textrm{HER}= \frac{\lvert\textrm{Hallucinated Tokens}\rvert}{\lvert\textrm{Reference Tokens}\rvert},
  \label{equation:eq1}
\end{align}
where the hallucinated tokens are identified as additional recognized tokens after the alignment process.
\\\textbf{Speech-augmented Language Models (SLMs) for ASR.} Instruction fine-tuning endows LLMs with the ability to transcribe speech by prompting them with acoustic features and task instructions \cite{shi2026qwen3, gong2024contextual}. As shown in Figure \ref{fig:fig2}, an archetype of SLMs comprises three components: a speech encoder, a connector network, and an LLM. The speech encoder, typically adapted from a speech foundation model (e.g., Whisper \cite{radford2023robust}), extracts high-level acoustic features $\mathrm{X}=(\mathbf{x}_{1},\mathbf{x}_{2},\cdots,\mathbf{x}_{T^{'}})$ from audio signals. To bridge the modality gap, the connector network compresses and projects $\mathrm{X}$ into a sequence of audio tokens $\mathrm{H}^{\mathrm{a}}=(\mathbf{h}_{1}^{\mathrm{a}},\mathbf{h}_{2}^{\mathrm{a}},\cdots,\mathbf{h}_{T}^{\mathrm{a}})$, with a sampling rate of $k$ (i.e., $T=T^{'}/k$). By prepending $\mathrm{H}^{\mathrm{a}}$ with the textual embeddings of an instruction sequence $\mathrm{W}^{\mathrm{i}}$ (i.e., task descriptions of ASR \cite{11434592}), the LLM then operates autoregressively over a reference transcription $\mathrm{W}=(w_{1},w_{2},\cdots,w_{N})$, modeling the likelihood of the next word emitted. The speech recognition process via an SLM can be expressed as: 
\begin{align}
  \mathcal{P}(\mathrm{W}|\mathrm{W}^{\mathrm{i}},\mathrm{X})=\prod_{n=1}^{N+1}\mathcal{P}(w_{n}|\mathrm{W}_{<n},\mathrm{W}^{\mathrm{i}},\mathrm{X}),
  \label{equation:eq1}
\end{align}
where $\mathrm{W}_{<n}$ is the sequence of previously generated tokens. The conditional probability of generating the $n$-th token $w_{n}$ is:
\begin{gather}
  \mathcal{P}(w_{n}|\mathrm{W}_{<n},\mathrm{W}^{\mathrm{i}},\mathrm{X}) = \mathrm{Softmax}(\mathrm{Lin}(\mathbf{h}_{n})), \label{equation:eq3} \\[0.5ex]
  \mathbf{h}_{n} = \mathrm{LLM}(\mathrm{H}_{<n}, \mathrm{H}^{\mathrm{i}},\mathrm{H}^{\mathrm{a}}), \label{equation:eq4}
\end{gather}
where the hidden representation $\mathbf{h}_n$ is derived from the LLM, conditioned on $\mathrm{H}^{\mathrm{a}}$, and the embeddings of instruction and previous outputs (viz. $\mathrm{H}^{\mathrm{i}}$ and $\mathrm{H}_{<n}$). In the training phase, we freeze the pretrained language model and integrate the LoRA module for efficient parameter adaptation.
\\\textbf{Causal Mediation Analysis of Hallucinations. }To diagnose hallucinations in SLMs, causal mediation analysis is leveraged to quantify the contributions of internal components in the backbone model (viz. MHA and MLP modules). We employ a knockout strategy through zero-ablation to selectively neutralize a target component $\theta_{c}$ in the SLM $\theta_{M}$ by zeroing the outputs of $\theta_{c}$ while preserving the remaining architecture and functionality intact \cite{wang2022interpretability, yang2025understanding}. For an ASR hypothesis $\hat{\mathrm{W}}$, the influence score $I_{\theta_{c}}$ of $\theta_{c}$ on a set of hallucinated words $\mathcal{H}$ is quantified as: 
\begin{equation}
\begin{split}
  I_{\theta_{c}} = \frac{1}{L} \sum_{t=1}^{T} & \mathbbm{1}\{\hat{w}_{t} \in \mathcal{H}\} \cdot \bigl| \mathcal{P}_{\theta_{M}}(\hat{w}_{t}|\hat{\mathrm{W}}_{<t}, \mathrm{W}^{\mathrm{i}}, \mathrm{X}) \\
  & - \mathcal{P}_{\theta_{M \setminus c}}(\hat{w}_{t}|\hat{\mathrm{W}}_{<t}, \mathrm{W}^{\mathrm{i}}, \mathrm{X}) \bigr|.
\end{split}
\label{equation:eq1}
\end{equation}
Here, $L$ is the total number of hallucinated words in $\mathcal{H}$, and $\theta_{M \setminus c}$ represents the model $\theta_{M}$ with component $\theta_{c}$ knocked out. $\mathbbm{1}\{\cdot\}$ is an indicator function, $\hat{w}_{t}$ denotes the token generated at time step $t$ based on the previously generated sequence $\hat{\mathrm{W}}_{<t}$. A high influence score indicates that $\theta_{c}$ has a strong contribution to hallucination generation.
\section{Mitigating Hallucination via AudioSLM}
\begin{figure*}[t]
  \noindent 
  \makebox[\textwidth][c]{ 
    \includegraphics[width=0.85\textwidth]{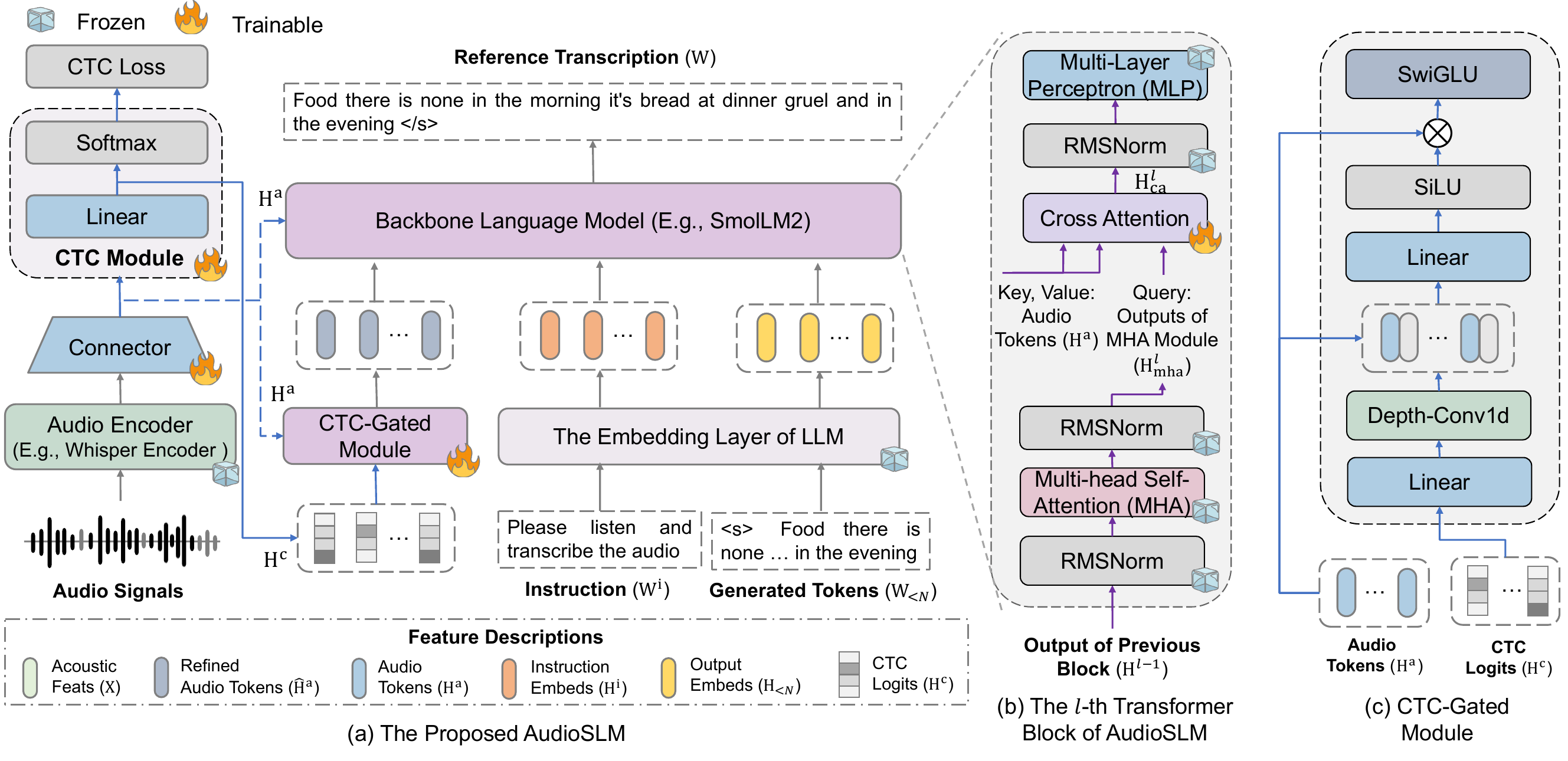}
  }
  \caption{The structure of AudioSLM for ASR to mitigate hallucination errors. As depicted in (a) the audio encoder and the connector first extract audio tokens from the audio signals, which are then transformed into alignment features via the CTC module and passed through the Gated-CTC module to provide the backbone language model with temporal cues; (b) the proposed Transformer block in AudioSLM, which facilitates fine-grained alignment across acoustic and linguistic modalities via a cross-attention layer; and (c) the Gated-CTC module, which leverages CTC logits to dynamically gate and refine the audio tokens.}
  \label{fig:fig3}
\end{figure*}

As shown in Figure \ref{fig:fig3}, AudioSLM enhances temporal cues and cross-modal alignment at both the input feature level and the neural architecture level. At input feature level, we derive a sequence of alignment features from the CTC module and incorporate them into the SLM via a CTC-Gated module. At neural architectural level, the Transformer blocks in the backbone language model are augmented with cross-attention layers to bolster the cross-modal alignment while preserving computational efficiency.
\begin{figure}
    \centering
    \includegraphics[width=1.0\linewidth]{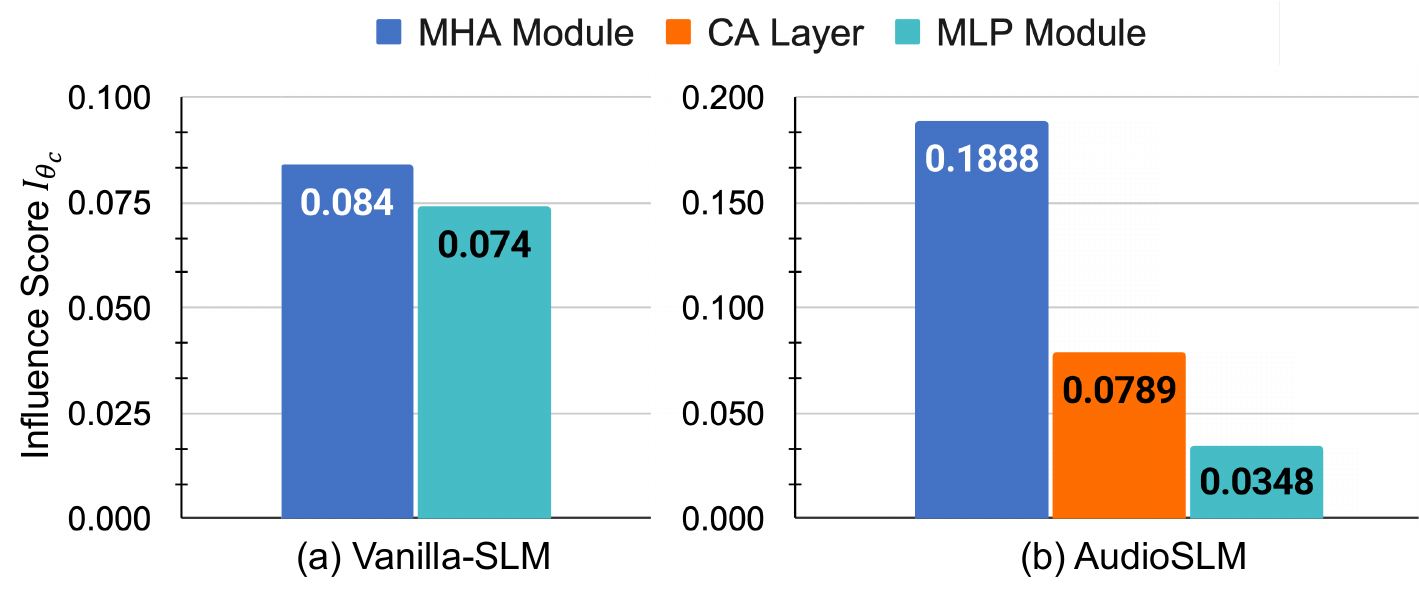}
    \caption{Comparison of influence scores for MHA and MLP modules, along with the cross-attention layer, for (a) Vanilla-SLM and (b) AudioSLM.}
    \label{fig:fig4}
\end{figure}
\\\textbf{Alignment Feature Integration for AudioSLM.} We build a CTC module on top of the connector to extract the logit vectors $\mathrm{H}^{\mathrm{c}}$ as alignment features, which are expected to render temporal cues via the CTC objective. Specifically, the CTC objective aligns audio tokens $\mathrm{H}^{\mathrm{a}}$ with the reference transcription $\mathrm{W}$ by marginalizing over all valid alignments and maximizing the conditional likelihood $\mathcal{P}(\mathrm{W}|\mathrm{H}^{\mathrm{a}})$ as: 
\begin{align}
  \mathcal{P}(\mathrm{W}|\mathrm{H}^{\mathrm{a}})=\sum_{\mathrm{A}\in\mathcal{B}^{-1}(\mathrm{W})}\mathcal{P}(\mathrm{A}|\mathrm{H}^{\mathrm{a}})\approx\sum_{\mathrm{A}\in\mathcal{B}^{-1}(\mathrm{W})}\prod_{t=1}^{T}\mathcal{P}(a_{t}|\mathrm{H}^{\mathrm{a}}),
  \label{equation:eq1}
\end{align}
where $\mathrm{A}=(a_{1},a_{2},\cdots,a_{T})$ denotes a valid alignment sequence drawn from the set $\mathcal{B}^{-1}(\mathrm{W})$, which contains all possible alignment paths compatible with $\mathrm{W}$. 
\begin{figure}[ht]
    \centering
    \includegraphics[width=1\linewidth]{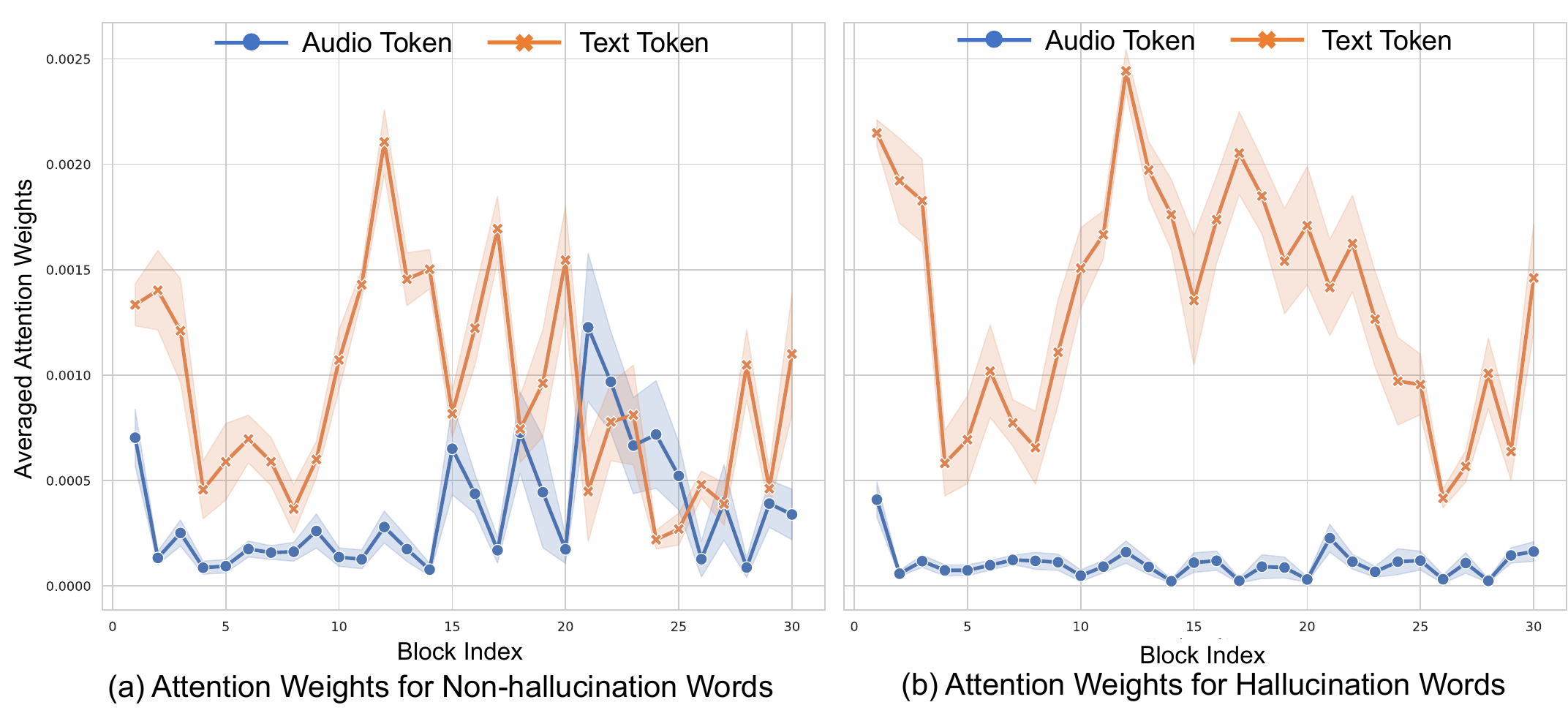}
    \caption{Averaged attention weights over audio and text tokens for hallucinated and non-hallucinated words in the Vanilla-SLM, where the shaded areas are the standard error of attention weights across all heads.}
    \label{fig:fig5}
\end{figure}
The token emission probability at time $t$ and the corresponding logit feature extraction are expressed as:
\begin{gather}
  \mathcal{P}(a_{t}|\mathrm{H}^{a})=\mathrm{Softmax}(\mathbf{h}_{t}^{\mathrm{c}}),
  \label{equation:eq4} \\[0.5ex]
  \mathbf{h}_{t}^{\mathrm{c}}=\mathrm{Lin}(\mathbf{h}_{t}^{\mathrm{a}}), \label{equation:eq5}
\end{gather}
where $\mathrm{Lin(\cdot)}$ denotes a linear transform projecting the hidden dimension into the vocabulary size of LLM. We then collect these logit features along the time axis to form the alignment features $\mathrm{H}^{\mathrm{c}}=(\mathbf{h}_{1}^{\mathrm{c}},\mathbf{h}_{2}^{\mathrm{c}},\cdots,\mathbf{h}_{T}^{\mathrm{c}})$. To fuse the audio tokens $\mathrm{H}^{\mathrm{a}}$ with alignment information in a fine-grained manner, the CTC-Gated module is employed. As shown in Figure \ref{fig:fig3}(c), CTC-Gated module produces refined audio tokens via element-wise gating and a Swish-Gated Linear Unit (SwiGLU) layer \cite{touvron2023llama}:
\begin{gather}
  \hat{\mathrm{H}}^{\mathrm{a}}=\mathrm{SwiGLU}(\mathrm{H}^{\mathrm{a}}+\mathrm{G}^{\mathrm{c}}\otimes\mathrm{H}^{\mathrm{a}}),
  \label{equation:eq5} \\[0.5ex]
  \mathrm{G}^{\mathrm{c}}=\sigma(\mathrm{Lin}_{2}([\mathrm{DC}(\mathrm{Lin}_{1}(\mathrm{H}^{\mathrm{c}})); \mathrm{H}^{\mathrm{a}}])), \label{equation:eq6}
\end{gather}
where $\otimes$ denotes the element-wise dot product. The gating vectors $\mathrm{G}^{\mathrm{c}}$ are computed by projecting $\mathrm{H}^{\mathrm{c}}$ from the vocabulary space to the hidden dimension via a linear projection $\mathrm{Lin}_{1}(\cdot)$. A depth-wise convolutional layer $\mathrm{DC}(\cdot)$ with a kernel size of $3\times1$ is then applied to capture local temporal dependencies. The resulting features are then concatenated with $\mathrm{H}^{\mathrm{a}}$ and transformed via a linear layer $\mathrm{Lin}_{2}(\cdot)$, followed by a sigmoid activation $\sigma(\cdot)$ to produce the gating vectors.
\\\textbf{Cross-Attention Layer for Transformer Block.} As illustrated in Figure \ref{fig:fig3}(b), the cross-attention layer is inserted between the MHA module and the MLP module. During training, we update only the parameters of the cross-attention layer, while keeping all others frozen. In the $l$-th Transformer block, the MHA module computes the contextualized representations $\mathrm{H}_{\mathrm{mha}}^{l}$ via the self-attention over the outputs of previous block $\mathrm{H}^{l-1}$. The cross-attention layer then takes $\mathrm{H}_{\mathrm{mha}}^{l}$ as queries and attends to the audio tokens $\mathrm{H}^{\mathrm{a}}$ for cross-modal alignment. The resulting representations, $\mathrm{H}_{\mathrm{ca}}^{l}$, are then processed by the MLP module to produce the block output $\mathrm{H}^{l}$: 
\begin{gather}
  \mathrm{H}_{\mathrm{mha}}^{l}=\mathrm{MHA}(\mathrm{H}^{l-1});\mathrm{H}_{\mathrm{ca}}^{l}=\mathrm{CA}(\mathrm{H}_{\mathrm{mha}}^{l}, \mathrm{H}^{\mathrm{a}}),
  \label{equation:eq6} \\[0.5ex]
  \mathrm{H}^{l}=\mathrm{MLP}(\mathrm{H}_{\mathrm{ca}}^{l}), \label{equation:eq7}
\end{gather}
where $\mathrm{CA}(\cdot)$ is cross-attention layer.
\section{Experiments}
\subsection{Experimental Settings}
\textbf{Dataset. }We conducted ASR experiments on the LibriSpeech corpus, a benchmark dataset for automatic speech recognition. We used the train-clean-100 subset for training, while evaluating on the dev-clean/other and test-clean/other sets.
\\\textbf{Implementation Details.} The audio encoder in AudioSLM was grafted from Whisper-large-v2, comprising 12 Transformer layers with 20 attention heads and a hidden dimension of 1280. The connector module consisted of a 3-layer convolutional neural network with a kernel size of 5, achieving a subsampling rate of 4. For the backbone model, we employed SmolLM-2-135M \cite{allal2025smollm2}, which consisted of 30 LLaMA blocks, each with 9 attention heads and a hidden size of 576. All models were trained for 30 epochs using a cosine learning rate scheduler with a maximum learning rate of 1e-4. During inference, the decoder beam size was set to 5. Our source code is publicly available at \texttt{https://github.com/bicheng1225/AudioSLM}.\\
\textbf{Comparative Methods.} We compare our proposed AudioSLM with the following top-of-the-line methods. As standard benchmarks in ASR system, CTC, RNN-T, and Joint CTC-Atten models leverage the Transformer architecture and are optimized through end-to-end objectives, including CTC, RNNT, and the hybrid encoder-decoder framework \cite{watanabe2017hybrid}. LLM-Guided Decoder \cite{higuchi2025harnessing} employs a large language model (i.e., LLaMA2-Chat-7B) as a front-end feature extractor, with an ASR decoder that refines hypotheses by conditioning on both textual representations and acoustic information. Vanilla-SLM is the variant of AudioSLM that adopts SmolLM2-135M as the backbone without any CTC-related components or cross-attention layers.
\begin{table}[t!]
  \caption{Detailed ASR Performance and hallucination error rates (HER\%) of SLMs on the Dev-clean.}
  \label{tab:tab1}
  \centering
  \resizebox{\columnwidth}{!}{
    \begin{tabular}{c|cccc|c}
      \toprule
      \textbf{Model} & \textbf{CORR}\%$\uparrow$ & \textbf{SUB}\%$\downarrow$ & \textbf{DEL}\%$\downarrow$ & \textbf{INS}\%$\downarrow$ & \textbf{HER}\%$\downarrow$ \\
      \midrule
      Vanilla-SLM & \textbf{94.58} & \textbf{4.78} & \textbf{0.63} & 3.69 & 52.01\\
      AudioSLM     & 94.16 & 5.20 & 0.64 & \textbf{0.74} & \textbf{8.69}\\
      \midrule
      -w/o CA     & \textbf{93.97} & \textbf{5.37} & 0.66 & 3.50 & \textbf{11.96}\\
      -w/o CTC-Gate& 93.93 & 5.55 & \textbf{0.52} & \textbf{1.52} & 14.67\\
      \bottomrule
    \end{tabular}
  }
  \vspace{2pt} 
  \begin{minipage}{\columnwidth}
    \footnotesize
  \end{minipage}
\end{table}
\subsection{Experimental Results}
\textbf{Causal Mediation Analysis of Hallucination in the SLM.} At the outset of experiments, we first compare the influence scores of the MHA and MLP to Vanilla-SLM and AudioSLM in Figure \ref{fig:fig4}(a), and then quantify the contribution of the introduced cross-attention (CA) layers to our model in Figure \ref{fig:fig4}(b). As shown in Figure \ref{fig:fig4}(a), both the MHA and MLP modules are critical for the generation of hallucinated words; however, the removal of MHA modules results to a larger impact than that of MLP modules. A consistent trend is observed in our AudioSLM, where MHA modules continue to exert the most profound influence, followed by the CA layers, both of which surpass the impact of the MLP modules in Figure \ref{fig:fig4}(b).
\\\textbf{Analyzing Hallucination Patterns through Attention Weights.} Building on the causal mediation analysis, we investigate the distributions of attention weights for hallucinated and non-hallucinated words to uncover distinct behavioral patterns. Figure \ref{fig:fig5} illustrates the attention weights in the MHA layers of the Vanilla-SLM, where weights assigned to audio and text tokens are averaged across heads and normalized by their respective sequence lengths. For non-hallucinated words, as shown in Figure \ref{fig:fig5}(a), attention weights exhibit an interleaving pattern across modalities, with text tokens dominating at shallower layers and audio tokens gaining prominence in deeper layers. In contrast, from Figure \ref{fig:fig5}(b), it is evident that the attention weights for hallucinated words are overwhelmingly skewed toward text tokens across all Transformer blocks, with negligible attention allocated to audio ones. This result aligns with the findings in computer vision \cite{leng2024mitigating, xie2024v}, where object hallucinations are attributed to an over-reliance on language priors, nearly setting aside visual information.
\\\textbf{Detailed ASR Performance and HER of SLMs on Dev-clean.} Table \ref{tab:tab1} summarizes the detailed ASR performance of the SLMs, reporting the correct recognition rate (CORR), substitution (SUB), deletion (DEL), insertion (INS), and hallucination error rates (HER). First, compared to Vanilla-SLM, AudioSLM significantly reduces the HER from 52.01\% to 8.69\%, accompanied by a marked improvement in insertion errors. However, these gains come at the cost of a slight degradation in other metrics (viz, CORR, SUB, DEL). Next, an ablation study is conducted to evaluate the contribution of each component in AudioSLM by separately removing the cross-attention layers (w/o CA) and the CTC-Gated module (w/o CTC-Gate). From the second part of Table \ref{tab:tab1}, it is evident that both proposed components are pivotal to alleviating hallucination, as the removal of either leads to an increase in HER compared to AudioSLM. Furthermore, these two components are complementary, as the CTC-Gate targets insertion errors and the CA layers are dedicated to suppressing hallucinations. These results suggest that temporal filtering via CTC-Gate effectively removes redundant frames, while strengthened cross-modal alignment via CA layers prevents the language model from relying solely on linguistic priors.
\begin{table}[t!]
  \caption{ASR Performance (WER\%) of AudioSLM and Comparative Methods on the LibriSpeech-100 Dataset.}
  \label{tab:tab2}
  \centering
  \resizebox{\columnwidth}{!}{
    \begin{tabular}{c|c|cc|cc} 
      \toprule
      \multirow{2}{*}{\textbf{Model}} & \multirow{2}{*}{\makecell[c]{\textbf{Used}\\\textbf{LLM}}} & \multicolumn{2}{c|}{\textbf{Dev WER}} & \multicolumn{2}{c}{\textbf{Test WER}} \\
      \cmidrule{3-6} 
      & & \textbf{Clean} & \textbf{Other} & \textbf{Clean} & \textbf{Other} \\
      \midrule
      CTC \cite{higuchi2022bert}   & -        & 11.20 & 21.40 & 11.40 & 22.00 \\
      RNN-T \cite{higuchi2022bert} & -        & 9.70  & 21.50 & 9.80  & 22.20 \\
      \makecell[c]{CTC-Atten \cite{higuchi2025harnessing}} & - & 7.20 & 17.50 & 7.50 & 18.00 \\
      \midrule
      \makecell[c]{LLM-Guided\\Decoder \cite{higuchi2025harnessing}} & LLaMA-7B-Chat & 6.20 & 16.50 & 6.70 & 16.90 \\
      Vanilla-SLM              & SmolLM2-135M &  9.11 & 13.75 & 10.82 & 13.93 \\
      \midrule 
      \multirow{3}{*}{AudioSLM} & SmolLM2-135M & 6.58 & 12.29 & 7.11 & 11.92 \\
                                       & SmolLM2-360M & 5.79 & 10.69 & 6.04& 11.07 \\
                                        & SmolLM2-1.7B & \textbf{4.69} & \textbf{8.75} & \textbf{4.71} &\textbf{10.21} \\
      \bottomrule
    \end{tabular}
  }
\end{table}
\\\textbf{Benchmarking SLMs on the Librispeech. }Table \ref{tab:tab2} compares AudioSLM with several strong baselines on the LibriSpeech test sets in terms of word error rate (WER). From this table, a consistent performance gain is observed, where AudioSLM (w/ SmolLM2-135M) outperforms the Vanilla-SLM and conventional architectures (e.g., CTC, RNN-T, and AED) under both clean and noisy conditions. In comparison with the LLM-Guided Decoder, AudioSLM (w/ SmolLM-135M) stands out on the dev-other and test-other, suggesting that our model exhibits greater robustness to acoustic variability and background noise. These findings are consistent with prior efforts on cold and deep fusion architectures, which sought to integrate language models into end-to-end ASR systems. \cite{sriram2017cold, tsunoo22_interspeech}. Finally, as the backbone model scales up, AudioSLM (w/ SmolLM2-1.7B) exhibits improved performance relative to both the LLM-Guided Decoder and its smaller-scale variants, validating the scalability and flexibility of the proposed architecture. Notably, in our empirical results, the HER(\%) for AudioSLM on Dev-clean rises from 8.69 to 10.64 and 12.76 as the backbone scales from 135M to 360M and 1.7B, respectively. This inverse scaling aligns with the observations in \cite{radford2023robust, peng2024owsm}, suggesting that larger speech foundation models are more susceptible to hallucinations due to their stronger language priors.
\section{Conclusion and Future Work}
In this paper, we have put forward AudioSLM for ASR to mitigate hallucination. Through causal and behavioral analysis, we identified that hallucinations primarily stem from attention weights biased toward textual representations. By incorporating alignment information and cross-attention layers, AudioSLM effectively reduced HER while maintaining competitive ASR performance. In future work, we plan to investigate training-free methods to suppress hallucinations via decoding strategies \cite{leng2024mitigating, zhao2024mitigating, adiga2025attention}.
\section{Acknowledgments}
This work was supported in part by Realtek Semiconductor Corporation under Grant Numbers 113KK01103 and 114KK01005. Any findings do not necessarily reflect those of the sponsors.
\section{Generative AI Use Disclosure}
The authors utilized Gemini-3-Flash to refine the wording clarity of this manuscript. We maintained full control over the research content and remain responsible for the accuracy and integrity of the experimental results and the presented figures.
\bibliographystyle{IEEEtran}
\bibliography{mybib}

\end{document}